\documentclass[aip,aipnum]{revtex4-1}

\usepackage{graphicx}
\usepackage{bm}
\usepackage[mathscr]{euscript}
\usepackage{verbatim}

\begin{document}

\title{Zonal flow generation and its feedback on turbulence production in drift wave turbulence}

\author{Andrey V. Pushkarev}

\author{Wouter J.T. Bos}
 \affiliation{LMFA, Ecole Centrale de Lyon, Ecully, France}

\author{Sergey V. Nazarenko}
 \affiliation{Mathematics Institute, University of Warwick, CV4 7AL, Coventry,
UK}

\date{\today}

\begin{abstract}
Plasma turbulence described by the Hasegawa-Wakatani equations has been simulated numerically for different models and values of the adiabaticity parameter $\mathscr{C}$. It is found that for low values  of $\mathscr{C}$ turbulence remains isotropic, zonal flows are not generated and there is no suppression of the meridional drift waves and of the particle transport. For high values of $\mathscr{C}$, turbulence evolves toward highly anisotropic states with a dominant contribution of the zonal sector to the kinetic energy. This anisotropic flow leads to a decrease of a turbulence production in the meridional sector and limits the particle transport across the mean isopycnal surfaces. This behavior allows to consider the Hasegawa-Wakatani equations a minimal PDE model which contains the drift-wave/zonal-flow feedback loop prototypical of the LH transition in plasma devices.
\end{abstract}

\maketitle
\section{\label{sec:Introduction}Introduction}
One of the major experimental discoveries in nuclear fusion research was the observation of a low-to-high (LH) transition in the plasma confinement characteristics \cite{Wagner1982}. This transition results in significantly reduced losses of  particles and energy from the bulk of the magnetically confined plasma and, therefore, improved conditions for nuclear fusion. Since this discovery, LH transitions have been routinely observed in a great number of modern tokamaks and stellarators, and the new designs like ITER rely on achieving H-mode operation in an essential way. The theoretical description of the LH transitions, and of nonlinear and turbulent states in fusion devices, is very challenging because of the great number of important physical parameters and scales of motion involved,  as well as a complex magnetic field geometry. Direct numerical simulations (DNS) of gyrokinetic Vlasov equations have become a popular tool for studying such fusion plasmas, which involves computing particle dynamics in a five-dimensional phase space (three space coordinates and two velocities) and, therefore, requires vast computing resources. Physical mechanisms to explain the LH transition have been suggested. One of these mechanisms is that small-scale turbulence, excited by a primary ({\it e.g.} ion-temperature driven) instability, drives a sheared zonal flow (ZF) via a nonlinear mechanism, through an anisotropic inverse cascade or a modulational instability. After this, the ZF acts to suppress  small-scale turbulence by shearing turbulent eddies or/and drift wave packets, thereby eliminating the cause of anomalously high transport and losses of plasma particles and energy.

Importantly, such possible scenario to explain the LH mechanism was achieved not by considering complicated realistic models but  by studying highly idealised and simplified models. More precisely, generation of ZF's by small-scale turbulence was predicted based on Charney-Hassegawa-Mima (CHM) equation \cite{Charney1948,hasegawa1978} very soon after this equation was introduced into plasma physics by Hassegawa and Mima in 1978~\cite{hasegawa1979nonlinear}, and even earlier in the geophysical literature~\cite{Rhines75}. The scenario of a feedback in which ZF's act onto small-scale turbulence via shearing and destroying of weak vortices was suggested by Biglari \textit{et al}. in 1990~\cite{biglari1990} using an even simpler model equation, which is essentially a 2D  incompressible neutral fluid description (equation (1) in~Ref.\onlinecite{biglari1990}). Probably the first instances where the two processes were described together as a negative feedback loop, turbulence generating ZF, followed by ZF suppressing turbulence, were in the papers by Balk \textit{et al}. 1990 \cite{Balk1990a, Balk1990b}. Balk \textit{et al}.  considered the limit of weak  wave-dominated  drift turbulence, whereas the picture of Biglari \textit{et al}. applies to strong eddy-dominated turbulence. In real situations, the degree of nonlinearity is typically moderate, \textit{i.e.} both waves and eddies are present simultaneously.
It is the relative importance of the anisotropic linear terms with respect to the isotropic nonlinear terms in the CHM equation which sets the anisotropy of the dynamics. If the linear terms are overpowered by the nonlinearity, the condensation of energy does not give rise to ZF's, but generate isotropic, round vortices.

Related, but more simplified models beyond CHM, are the modified CHM~\cite{Hammett1992}, Hasegawa-Wakatani (HW)~\cite{Hasegawa1983} model  and modified Hasegawa-Wakatani model~\cite{numata2007}. The HW model is given by equations (\ref{eq:HW1}) and  (\ref{eq:HW2}) below. The term ``modified" in reference to both the CHM and HW models means that the zonal-averaged component is subtracted from the electric potential to account for absence of the Boltzmann response mechanism for the mode which has no dependence in the direction parallel to the magnetic field.

Quantitative investigations of the LH transition physics are presently carried out, using realistic modelling, such as gyrokinetic simulations, drawing inspiration from the qualitative results obtained by these idealised models. However, the understanding of the dynamics generated by these idealised models remains incomplete. It was only recently that the scenario of the drift-wave/ZF feedback loop proposed theoretically in 1990 for the CHM model  was confirmed and validated by Direct Numerical Simulations (DNS) of the CHM equation by Connaughton \textit{et al}~\cite{Nazarenko2011}. In their work, the system was forced and damped by adding a linear term on the  right-hand side of the CHM equation which would mimic a typical shape of a relevant plasma instability near the Larmor radius scales and dissipation at smaller scales. A  drift-wave/ZF feedback loop was also seen in DNS  of the modified HW model by Numata \textit{et al}~\cite{numata2007}. The two dimensional simplification of the HW equations involves a coupling parameter generally called the adiabaticity. In one limit of this adiabaticity parameter the HW model becomes CHM model and in another limit it becomes the 2D Euler equation for an incompressible neutral fluid. The HW model contains more physics than CHM in that it contains turbulence forcing in the form of a (drift dissipative) instability and it predicts a non-zero turbulent transport - both effects are absent in CHM. On the other hand, it was claimed in Numata \textit{et al}.~\cite{numata2007} that the original HW model (without modification) does not predict formation of ZF's. This claim appears to be at odds with the CHM results of Connaughton \textit{et al}., considering the fact that HW model has CHM as a limiting case.

In the present work we will perform DNS  of the  HW model (without modification) aimed at checking realisability of  the drift-wave/ZF feedback scenario proposed by Balk \textit{et al}. in~1990 \cite{Balk1990a,Balk1990b} and numerically observed by Connaughton \textit{et al}~\cite{Nazarenko2011}. This will be a step forward with respect to the CHM simulations because the instability forcing is naturally present in the HW model and there is no need to add it artificially as it was done for CHM. We will vary over a wide range of the coupling parameter of the HW model including its large values which bring HW close to the CHM limit. We will see that  the ZF generation and turbulence suppression are indeed observed for such values of the coupling parameter, whereas for its smaller values these effects are lost.

\section{\label{sec:ModelHW}Physical model}
The model we will consider is based on the HW equations~\cite{Hasegawa1983}:
\begin{eqnarray}\label{eq:HW1}
 \left(\frac{\partial}{\partial t} -
  \nabla\psi\times\bf{z}\cdot\nabla\right)\nabla^{2}\psi =
  \mathscr{C}\left(\psi-n\right) - \nu\nabla^{4}(\nabla^{2}\psi),
\end{eqnarray}
\begin{eqnarray}\label{eq:HW2}
 \left(\frac{\partial}{\partial t} -
  \nabla\psi\times\bf{z}\cdot\nabla\right)(n+ln(n_{0})) =
  \mathscr{C}\left(\psi-n\right) - \nu\nabla^{4}n,
\end{eqnarray}
where $\psi$ is  the electrostatic potential, $n$ is the
density fluctuation. The variables in Eq.~(\ref{eq:HW1}) and Eq.~(\ref{eq:HW2})
have been normalized as follows,
\begin{eqnarray*}
x/\rho_{s}\rightarrow x~,~~\omega_{ci}t\rightarrow t~,~~ e\psi/T_{e}\rightarrow
\psi~,~~ n_{1}/n_{0}\rightarrow n,
\end{eqnarray*}
where $\rho_{s}=\sqrt{T_{e}/m}\omega_{ci}^{-1}$ is the ion gyroradius, $n_{0}$ and $n_{1}$ are the mean and the fluctuating part of the density, $e, m$ and $T_{e}$ are the electron charge, mass and temperature respectively and $\omega_{ci}$ is the ion cyclotron frequency. $\mathscr{C}$ is the adiabaticity parameter which we will discuss below. The last terms in Eq.~(\ref{eq:HW1}) and Eq.~(\ref{eq:HW2}) are $4^{th}$-order hyperviscous terms that mimic small scale damping.

The physical setting of the HW model may be considered as a simplification of the edge region of a tokamak plasma in the presence of a nonuniform background density $n_{0}=n_{0}(x)$ and in a constant equilibrium magnetic field $\textbf{B}=B_{0}{\bf e}_{z}$, where ${\bf e}_{z}$ is a unit vector in the $z$-direction. The assumption of  cold ions and isothermal electrons allows one to find Ohm's law for the parallel electron motion:
\begin{eqnarray}
\eta_{\parallel}J_{\parallel}=-nev_{\parallel}=
E_{\parallel}+\frac{1}{ne}\nabla_{\parallel}p=\frac{T_e}{ne}\nabla_{\parallel}n-
\nabla_{\parallel}\psi,
\end{eqnarray}
where $\eta_{\parallel}$ is the parallel resistivity, $E_{\parallel}$ is the parallel electric field, $v_{\parallel}$ is the parallel electron velocity and $p$ is the electron pressure. This relation gives the coupling of Eq.~(\ref{eq:HW1}) and Eq.~(\ref{eq:HW2}) through the adiabaticity operator $\mathscr{C}=-T_{e}/(n_{0}\eta\omega_{ci}e^{2})\partial^{2}/\partial z^{2}$. We will show in the following that the HW model describes the growth of small initial perturbations due to the linear drift-dissipative instability leading to drift wave turbulence evolving to generate ZF via an anisotropic inverse cascade mechanism followed by suppression of drift wave turbulence by ZF shear.

The role of the dissipation terms in the equations~(\ref{eq:HW1})~and~(\ref{eq:HW2}) is to ensure the possibility of a steady state and to prevent a spurious accumulation of energy near the smallest resolved scales. We chose the dissipation  terms proportional to $\nu k^{4}$, but the qualitative picture is expected to be largely insensitive to the particular choice of the dissipation function.

The important, relevant quantity in fusion research is the particle flux in the $x$ direction due to the  fluctuations,
\begin{eqnarray}\label{eq:particleFlux}
\Gamma_{n}=\kappa\int n\frac{\partial \psi}{\partial y}dV,
\end{eqnarray}
where $\kappa=\rho_{s}|\nabla ln(n_{0})|$ is the normalized density
gradient. Another quantity that we will monitor is the total energy,
\begin{eqnarray}
E^{T}=E + E^{n} =\frac{1}{2}\int(|\nabla\psi|^{2}+n^{2})dV.
\end{eqnarray}
We will be interested in particular in the velocity field, the ZF's and their influence on turbulent fluctuations. We therefore focus on the kinetic energy. Since one of the main subjects of the present work is the investigation of the ZF generation, we will quantify the energy contained in these ZF's by separating the energy into the kinetic energy $E_{z}$ contained in a zonal sector, defined as $|k_{x}|>|k_{y}|$, and the kinetic energy $E_{m}$ contained in a meridional sector, $|k_{x}|\leq|k_{y}|$.

\begin{figure}[h]
\center{
\begin{tabular}{cc}
  \includegraphics[width=5cm]
{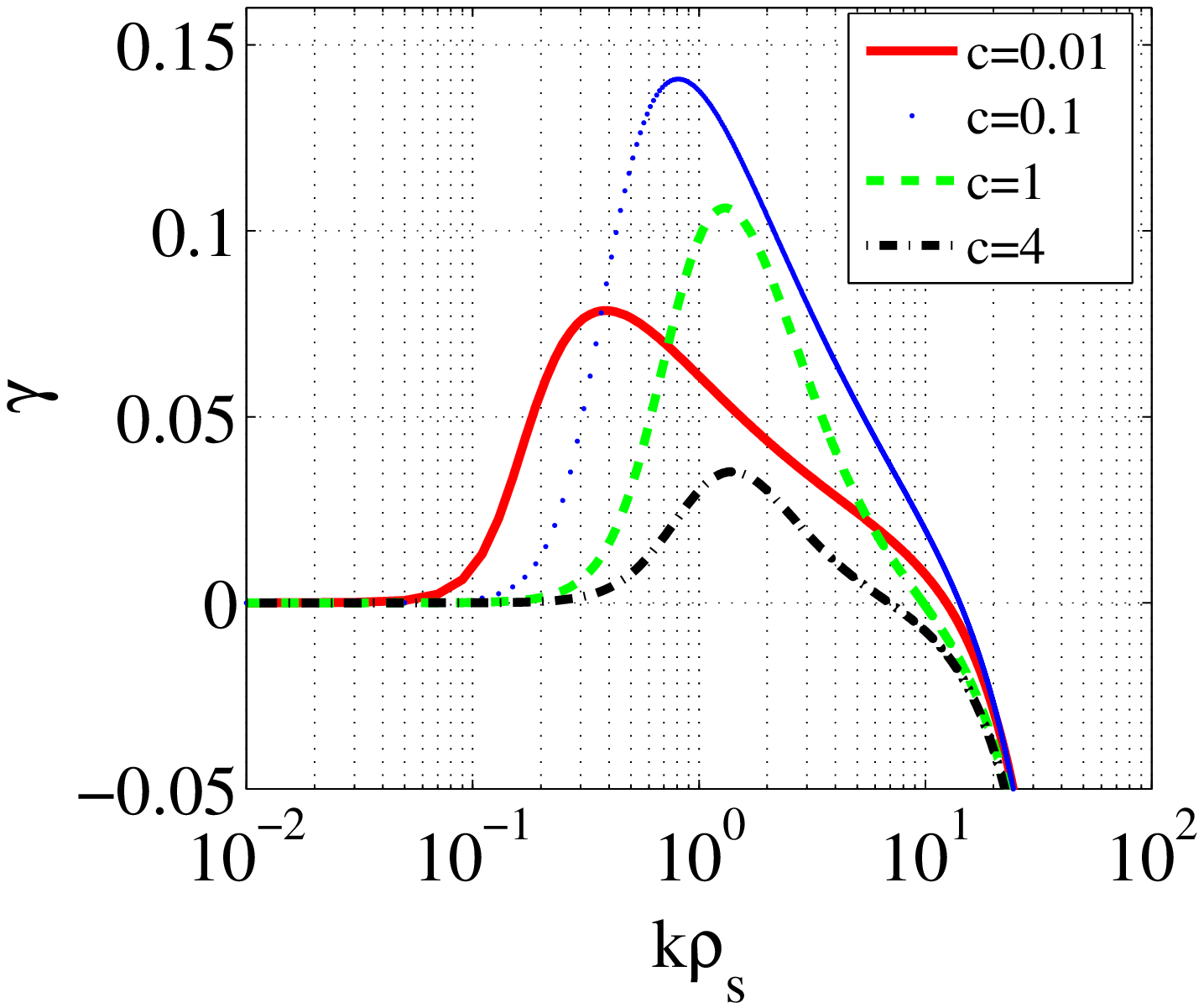} &
  \includegraphics[width=5cm]
{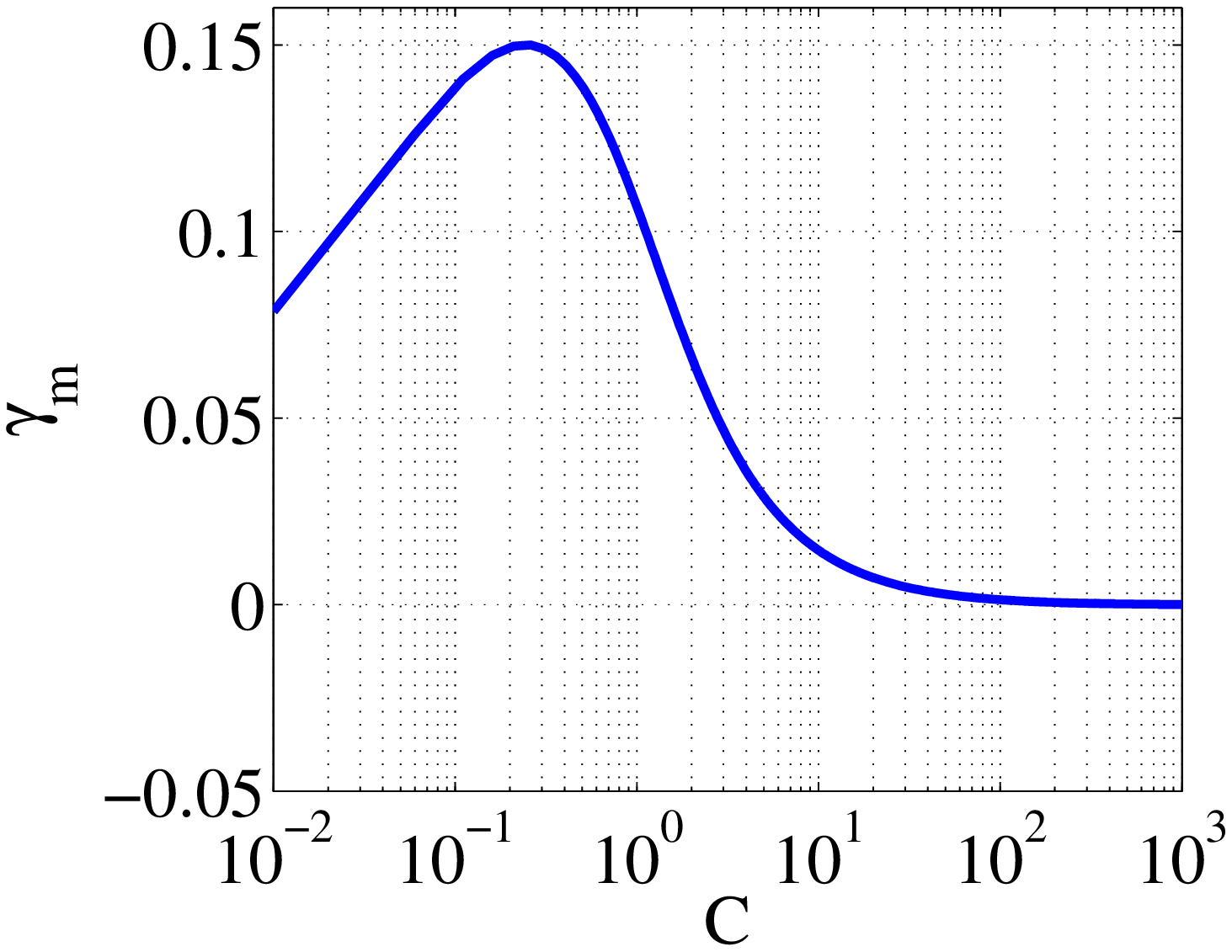} \\
    (a) & (b)
\end{tabular}
}
\caption{(a) Linear growth rate $\gamma$ for $k_{x}=0$, $\nu=10^{-5}$ and different values of
$\mathscr{C}$:  $\mathscr{C}=0.01$ (solid line),
$\mathscr{C}=0.1$ (dashed line), $\mathscr{C}=1$ (dotted line), $\mathscr{C}=4$
(dash-dot line). (b) Maximum linear growth rate $\gamma_{max}$ for $k_{x}=0$ as a
function of $\mathscr{C}$, for $\nu=10^{-5}$.}
\label{ris:stableAnalisHW}
\end{figure}

Different possibilities to determine the adiabaticity parameter will now be discussed. The first one is based on the linear stability analysis. In order to understand the dependence of the HW model on its constituent parameters it is useful to solve the linearised system. Linearisation of the Eq.~(\ref{eq:HW1}) and Eq.~(\ref{eq:HW2}) around the zero equilibrium ($\psi=0$ and $n=0$) and considering a plane wave solution, $\psi(k,t)\sim\psi_{0}e^{i(\bm k\cdot \bm x -\omega t)}$ and $n(k,t)\sim n_{0}e^{i(\bm k\cdot \bm x -\omega t)}$, yields  the dispersion relation for a resistive drift wave:
\begin{eqnarray}\label{eq:dispRelationHW}
\omega^{2}+i\omega(b+2\nu k^{2})-ib\omega_{*}
-\mathscr{C}\nu k^{2}(1+k^{2})-\nu^{2}k^{8}=0,
\end{eqnarray}
where $\omega_{*}=k_{y} \kappa/(1+k^{2})$ is the drift frequency,
$b=\mathscr{C}(1+k^{2})/k^{2}$
and $k^{2}=k^{2}_{x}+k^{2}_{y}$.
Let us introduce the
real frequency $\omega_{R}$ and the  growth rate $\gamma$ as
\begin{eqnarray}
\omega=\omega_{R}+i\gamma
\end{eqnarray}
The dispersion relation (\ref{eq:dispRelationHW}) has two solutions, a stable one, with $\gamma_{max} = \max \gamma({\bf k}) >0$, and an unstable one,
with $\gamma({\bf k}) \le 0$.
The Fig.~\ref{ris:stableAnalisHW}(a) shows the behaviour of $\gamma$ for $k_{x}=0$ and $\nu=10^{-5}$ for the unstable mode as a function of the wavenumber. The behaviour of $\gamma_{max} $  as a function of $\mathscr{C}$ is shown in Fig.~\ref{ris:stableAnalisHW}(b).

For the inviscid case, if $\nu$ is ignored, the solution  of Eq.~(\ref{eq:dispRelationHW}) is
\begin{eqnarray}
\omega=\frac{1}{2}[-ib\pm ib(1-4i\omega_{*}/b)^{0.5}].
\end{eqnarray}
The maximum growth rate corresponds to $b\simeq 4\omega_{*}$  and
\begin{eqnarray}\label{eq:c1hw}
\mathscr{C}=\frac{4k^{2}k_{y}\kappa}{(1+k^{2})^{2}}.
\end{eqnarray}

The adiabaticity operator $\mathscr{C}=-T_{e}/(n_{0}\eta\omega_{ci}e^{2})\partial^{2}/\partial z^{2}$ in Fourier space becomes an adiabaticity parameter via replacement  $\partial^{2}/\partial z^{2} \to -k_{z}^{2}$, where $k_{z}$ is a wavenumber characteristic of the fluctuations of the drift waves along the field lines in the toroidal direction.
Recall that we assume  the fluctuation length scale to satisfy the drift ordering, $k_{\parallel}\ll k_{\perp}$.
It is natural to
assume that the system selects $k_{z}$ which corresponds to the fastest growing wave mode. In this case
one should choose the parallel wavenumber $k_{z}$ which satisfies Eq.~(\ref{eq:c1hw}) for each fixed value of perpendicular wavenumber ${\bf k}_{\perp}=(k_{x},k_{y})$. This approach is valid provided that the plasma remains collisional for this value of $k_{z}$.  Note that according to  Eq.~(\ref{eq:c1hw}) such a choice gives $\mathscr{C}=0$ for the modes with $k_{y}=0$.

Another common approach is to define the parameter $\mathscr{C}$ simply as a constant. This approach makes sense if the maximum growth rate correspond to values of  $k_{z}$ which are smaller than the ones allowed by the finite system, i.e. $k_{z \, min} = 1/R$, where $R$ is the bigger tokamak radius. In this case the HW model has two limits: adiabatic weak collisional limit ($\mathscr{C}\rightarrow\infty$) where the system reduces to the CHM equation, and the hydrodynamical limit ($\mathscr{C}\rightarrow0$) where the system of Eqs.~(\ref{eq:HW1}) and~(\ref{eq:HW2}) reduces to the system of  Navier-Stokes equations and an equation for a  passive scalar mixing.

In our simulation we will try and compare both approaches: choosing constant $\mathscr{C}$ and chosing $\mathscr{C}$ selected by the maximum growth condition (\ref{eq:c1hw}).

\section{\label{sec:NumericalMethod}Numerical method}
Numerical simulations were performed using a pseudo-spectral Fourier code on a square box with periodic boundary conditions. The number of the modes varied from $256^{2}$ (with the lowest wavenumber $\triangle k=0.042$ and the size of the box $L_{x}=L_{y}=150$) to $1024^{2}$ (with the lowest wavenumber $\triangle k=0.002$ and the box size  $L_{x}=L_{y}=300$), the viscosity coefficient  was taken $\nu=0.0005$ and $\nu=0.00005$, respectively. The time integration was done by the fourth-order Runge-Kutta method. The integration time step was taken to be $\triangle t=5\cdot10^{-4}$ and $\triangle t=10^{-4}$.

\section{\label{sec:Results}Results}
In the following we will present our results for the evolution of the HW turbulence for different choices of the form and the values of the adiabaticity parameter.
\paragraph{Constant adiabaticity}
In this section we will consider the case where the adiabaticity parameter is taken  a constant. We present the results obtained with different values of $\mathscr{C}$ corresponding to the hydrodynamic regime ($\mathscr{C}\rightarrow0$, strongly collisional limit), adiabatic regime ($\mathscr{C}\rightarrow\infty$, weakly collisional limit) and transition regime ($\mathscr{C} \simeq 1$). The simulation parameters are presented in Table~\ref{tab:table1}.
\begin{table}[h]
\caption{The simulation parameters.}
\begin{ruledtabular}
\begin{tabular}{cccc}
 $\mathscr{C}$ & 0.01 & 1 & 40 \\
\hline
$L_{x}\times L_{y}$& $300\times300$ & $300\times300$ &  $150\times150$ \\
$N_{x}\times N_{y}$& $1024^{2}$ & $1024^{2}$ &  $256^{2}$ \\
$\rho_{s}$& 0.02 & 0.02 & 0.04 \\
$\kappa$& 0.3491 & 0.3491 & 0.0418 \\
\end{tabular}
\end{ruledtabular}
\label{tab:table1}
\end{table}

\begin{figure}[t]
\begin{tabular}{ccc}
  \includegraphics[width=5.5cm]
{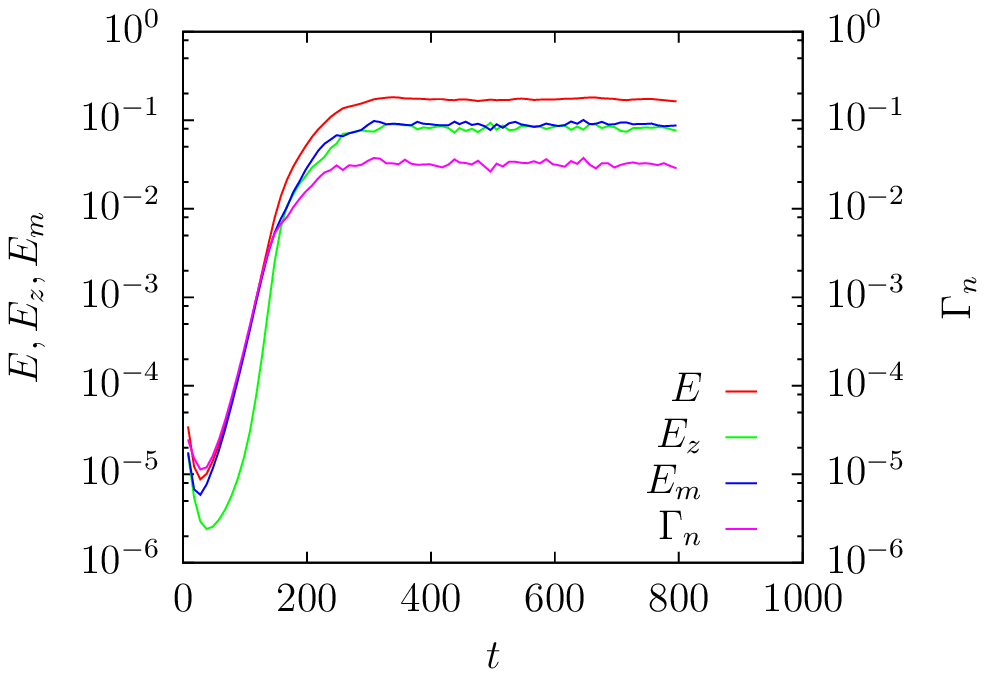} &
  \includegraphics[width=5.5cm]
{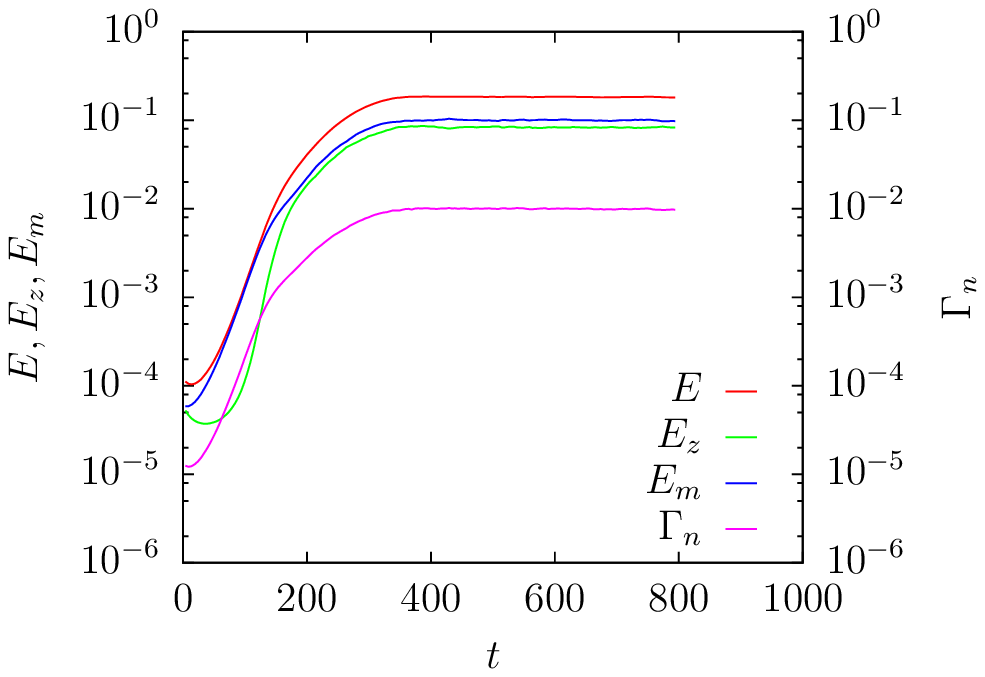} &
  \includegraphics[width=5.5cm]
{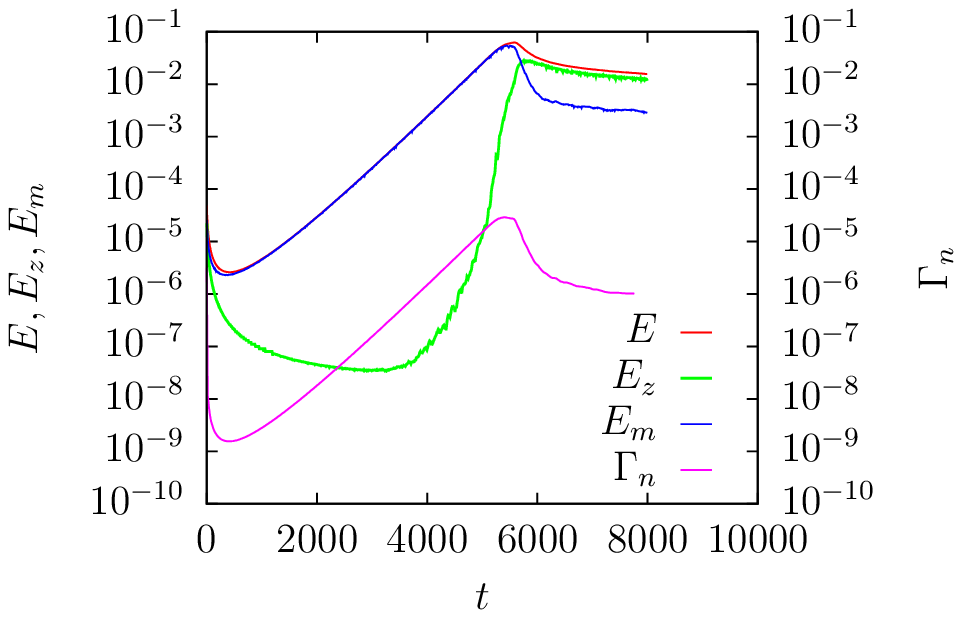} \\
    (a) & (b) & (c)
\end{tabular}
\caption{Evolution of the total kinetic energy, energy contained in the zonal and the
meridional sectors. (a) $\mathscr{C}=0.01$, (b) $\mathscr{C}=1$ and (c)
$\mathscr{C}=40$.}
\label{ris:energy_1024_difC}
\end{figure}

Fig.~\ref{ris:energy_1024_difC} shows the typical time evolution of the total kinetic energy, kinetic energy contained in the zonal sector $|k_{x}|>|k_{y}|$, kinetic energy contained in the meridional sector $|k_{x}|\leq|k_{y}|$ and the particle flux $\Gamma_{n}$ (see Eq.~(\ref{eq:particleFlux})), for the adiabaticity parameter values $\mathscr{C}=0.01,1$ and 40, respectively. From Fig.~\ref{ris:energy_1024_difC} we see that the small initial perturbations grow in the initial phase. In this phase the amplitudes of the drift waves grow. Then, these drift waves start to interact nonlinearly. For the case $\mathscr{C}=0.01$ and $\mathscr{C}=1$ the resulting saturated state seems close to isotropic as far as can be judged from the close balance between $E_{z}$ and $E_{m}$. For the simulation with $\mathscr{C}=40$ it is observed that the meridional energy strongly dominates until $t\approx4000$. After this, the zonal energy rapidly increases and becomes dominant for $t>6000$. This picture is in agreement with the scenario proposed in Connaughton \textit{et al}.~\cite{Nazarenko2011} for the CHM system. For the different values of $\mathscr{C}$ we can observe distinct types of  behaviour in the evolution of the kinetic energy. The initial phase  always agrees with the linear stability analysis (section~\ref{sec:ModelHW}). The speed at which the system enters to the saturated state is strongly dependent on $\mathscr{C}$.
The slowness of the transition of the system to a saturated level has limited the maximum value of the adiabaticity parameter to $\mathscr{C}=40$ and the maximum number of modes for such $\mathscr{C}$ to $256^2 $. This slowness can be understood from the linear growth rate dependence which decreases rapidly with $\mathscr{C}$, see Fig.~\ref{ris:stableAnalisHW}(b).
\begin{figure}[h!]
\center{
\begin{tabular}{cc}
  \includegraphics[width=5.5cm]
{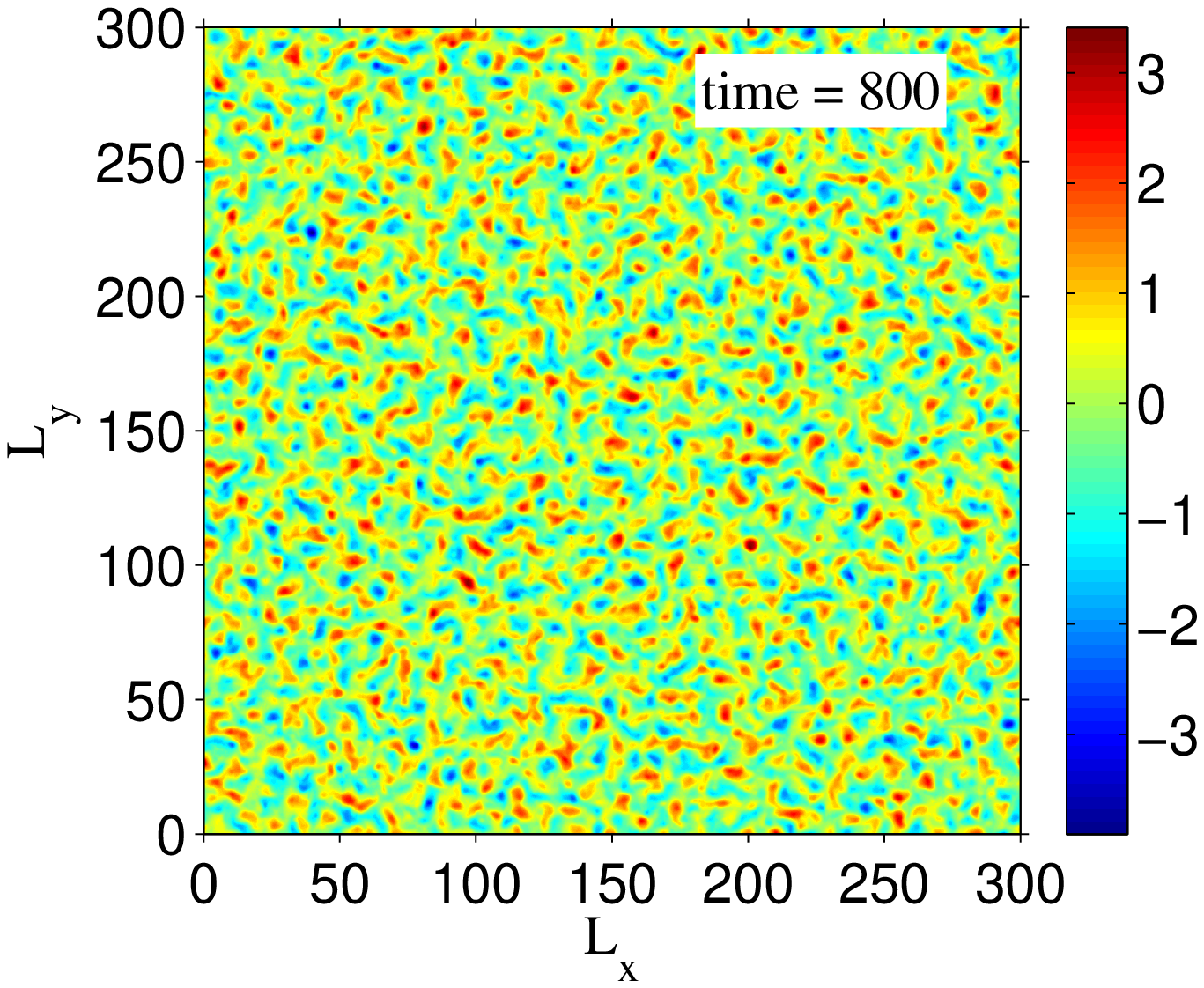} &
  \includegraphics[width=5.5cm]
{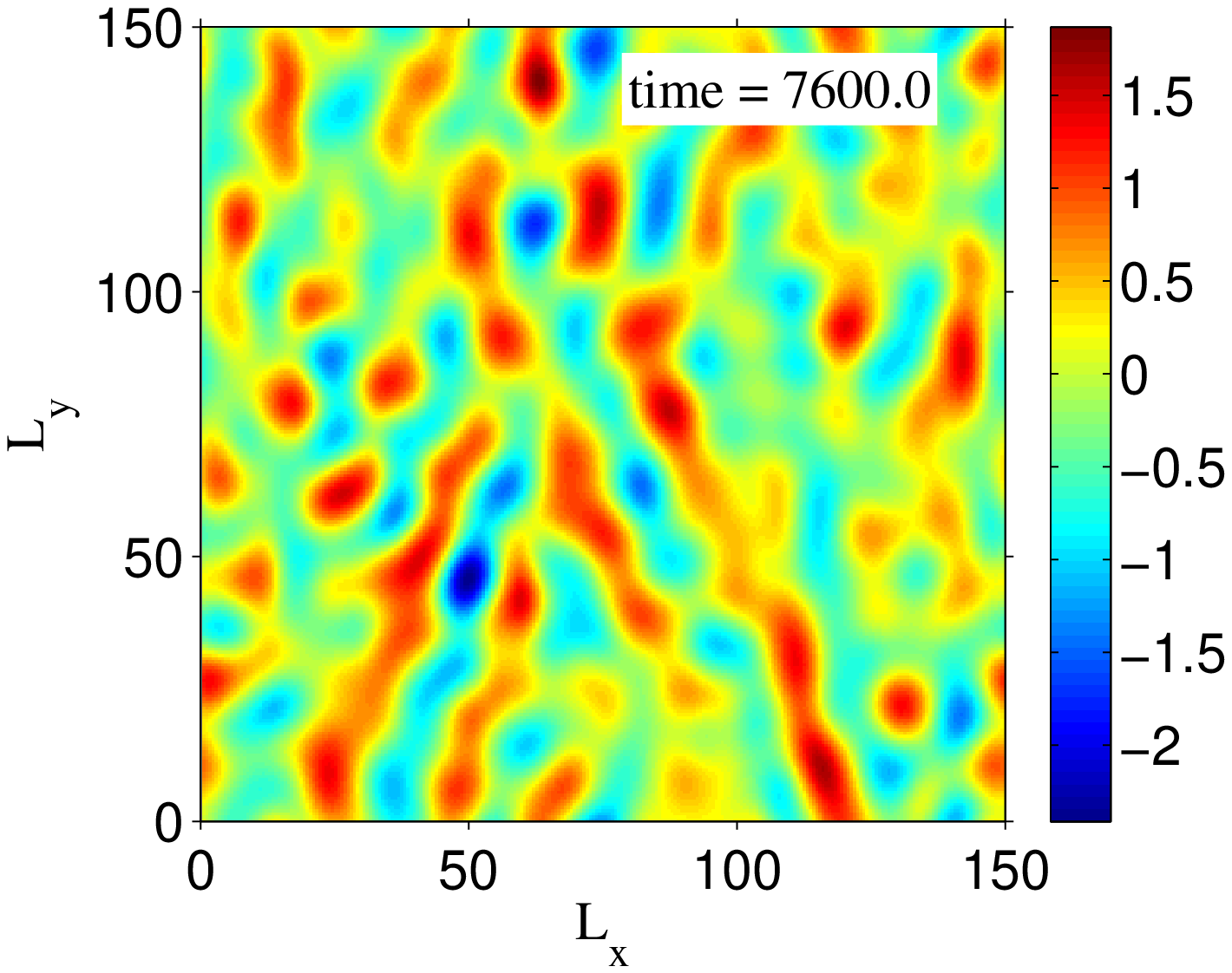} \\
    (a) & (b)
\end{tabular}
}
\caption{Fields of the stream functions. a) $\mathscr{C}=1$ $t=800$,
b) $\mathscr{C}=40$ $t=7600$.}
\label{ris:fld_ps_c1_c40}
\end{figure}
For the value $\mathscr{C}=0.01$ and $\mathscr{C}=1$ we observe monotonous growth of the zonal, meridional and the total energies, as well as the  the particle flux $\Gamma_{n}$ -- until these quantities reach saturation.
We see that for $\mathscr{C}=40$ the initial growth of the meridional energy and the particle flux $\Gamma_{n}$ is followed by a significant (between one and two orders of magnitude) suppression of their levels at the later stages. This is precisely the type of behavior previously observed in the CHM turbulence~(Ref.\onlinecite{Nazarenko2011}), and which corresponds to LH-type transport and drift-wave suppression.
Recall that  $\Gamma_{n}$ is the particle flux in  the \textit{x}-direction which corresponds to the radial direction of the physical system that we model, the edge region of the  tokamak plasma.

Fig.~\ref{ris:fld_ps_c1_c40}(a) and Fig.~\ref{ris:fld_ps_c1_c40}(b) show instantaneous visualisations of the electrostatic potential $\psi$ for values $\mathscr{C}=1$ and $\mathscr{C}=40$, respectively. The structure of $\psi$ is strongly dependent on the regime: for low values of $\mathscr{C}$ the structure of $\psi$ is isotropic, whereas for the high values of $\mathscr{C}$ the structure of $\psi$ is anisotropic and characterised by formation of large structures elongated in the zonal direction.

\begin{figure}[h]
\center{
\begin{tabular}{cc}
  \includegraphics[width=5.5cm]
{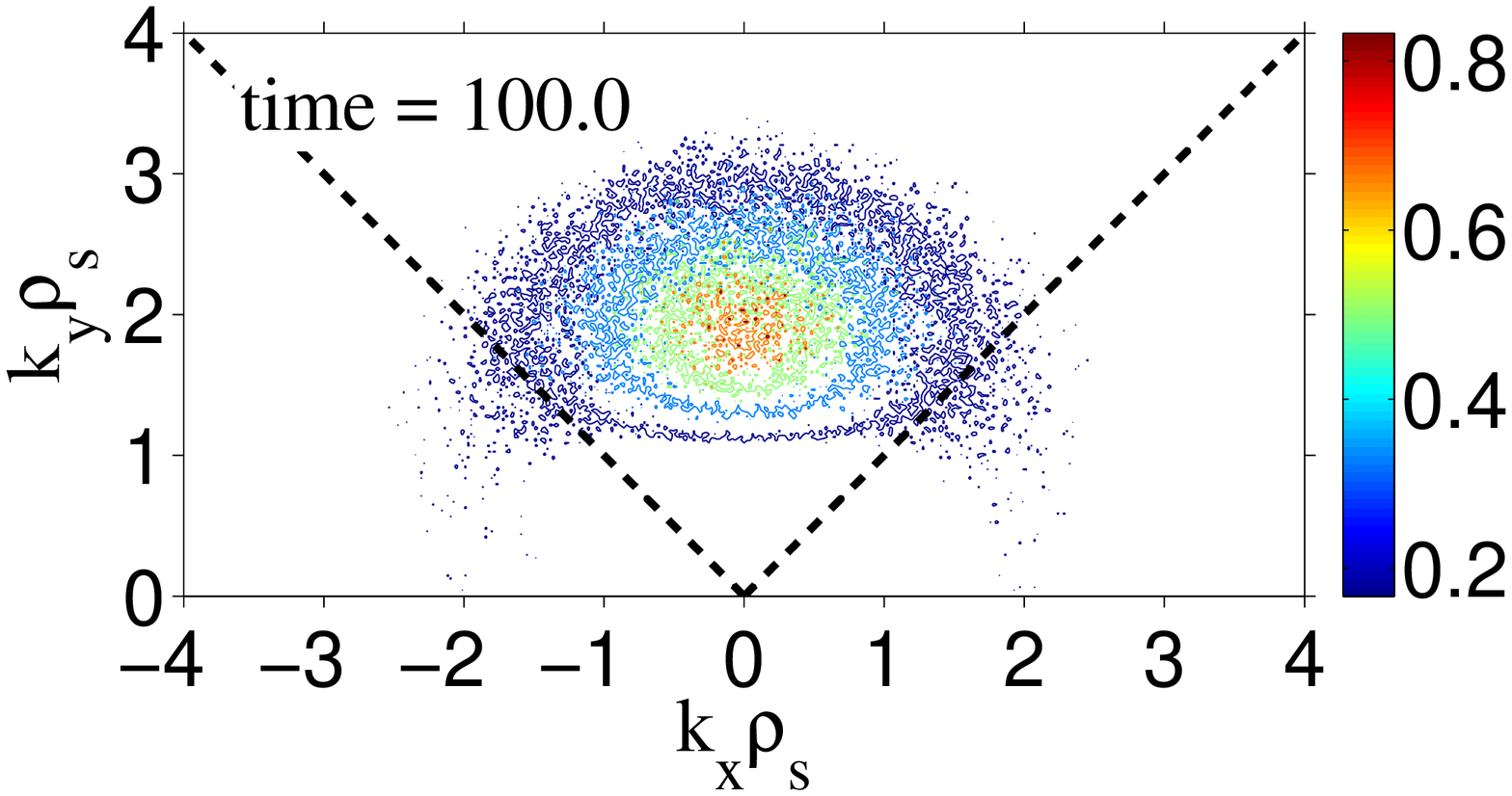} &
  \includegraphics[width=5.5cm]
{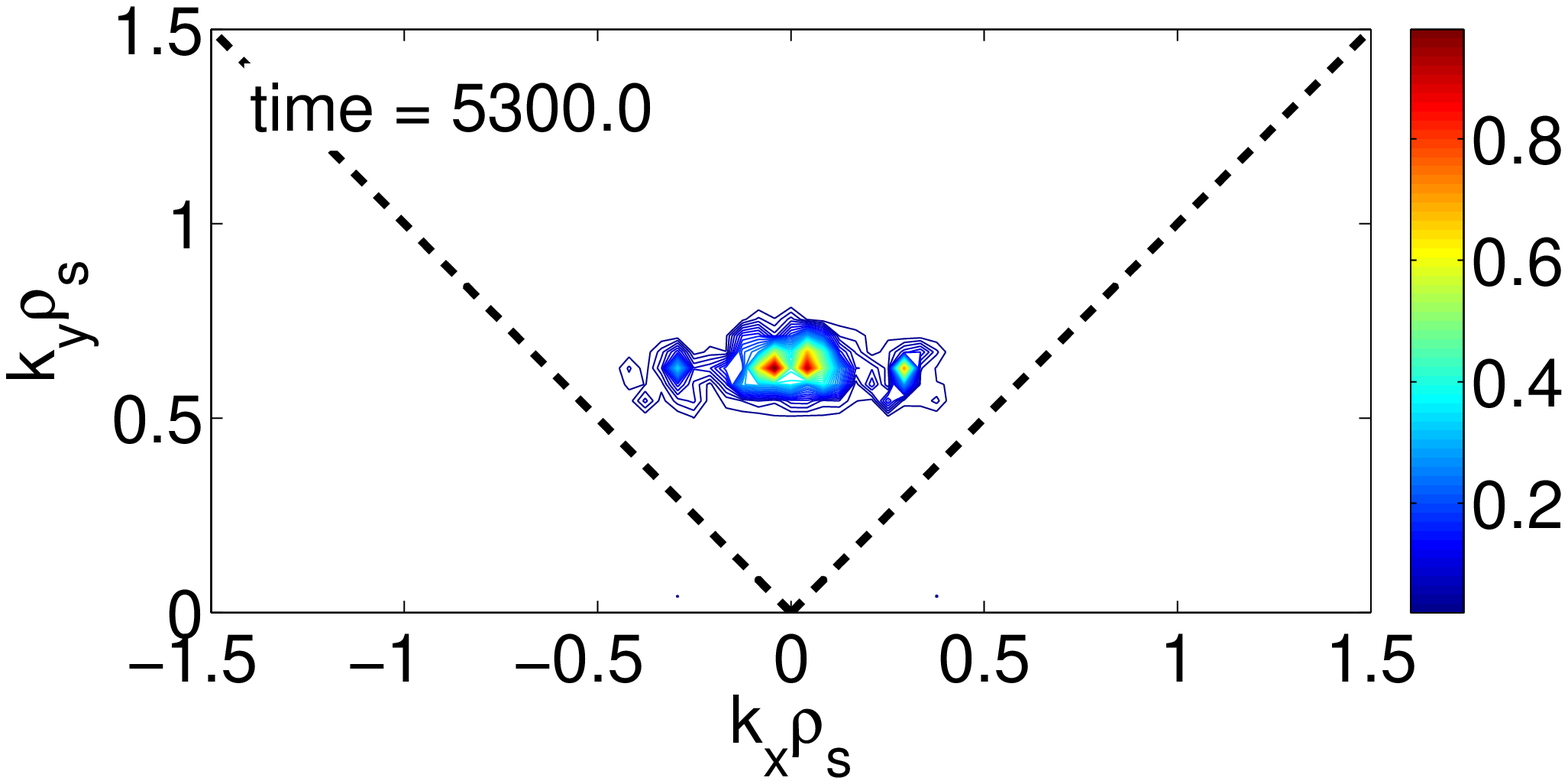} \\
    a) & b)  \\
  \includegraphics[width=5.5cm]
{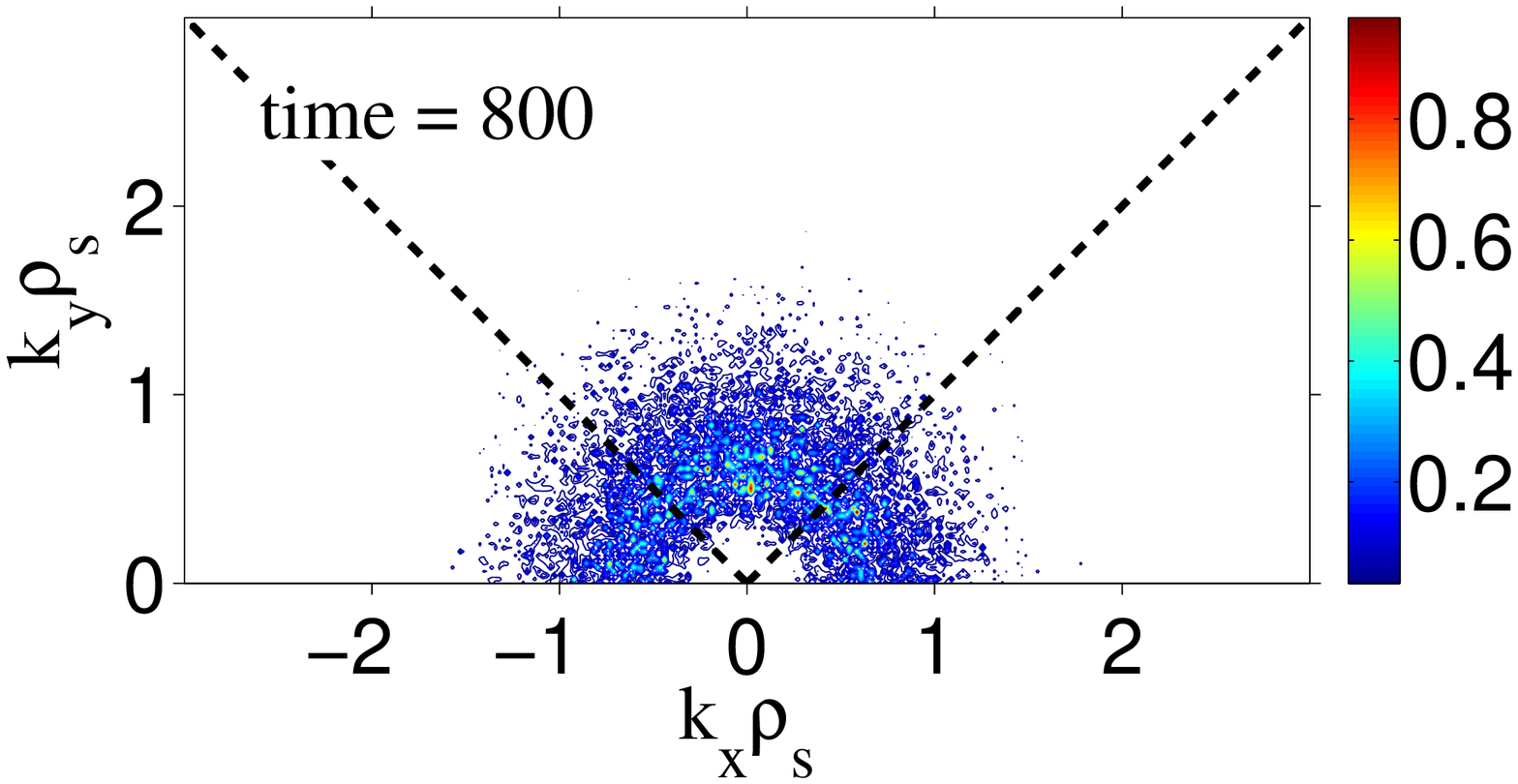} &
  \includegraphics[width=5.5cm]
{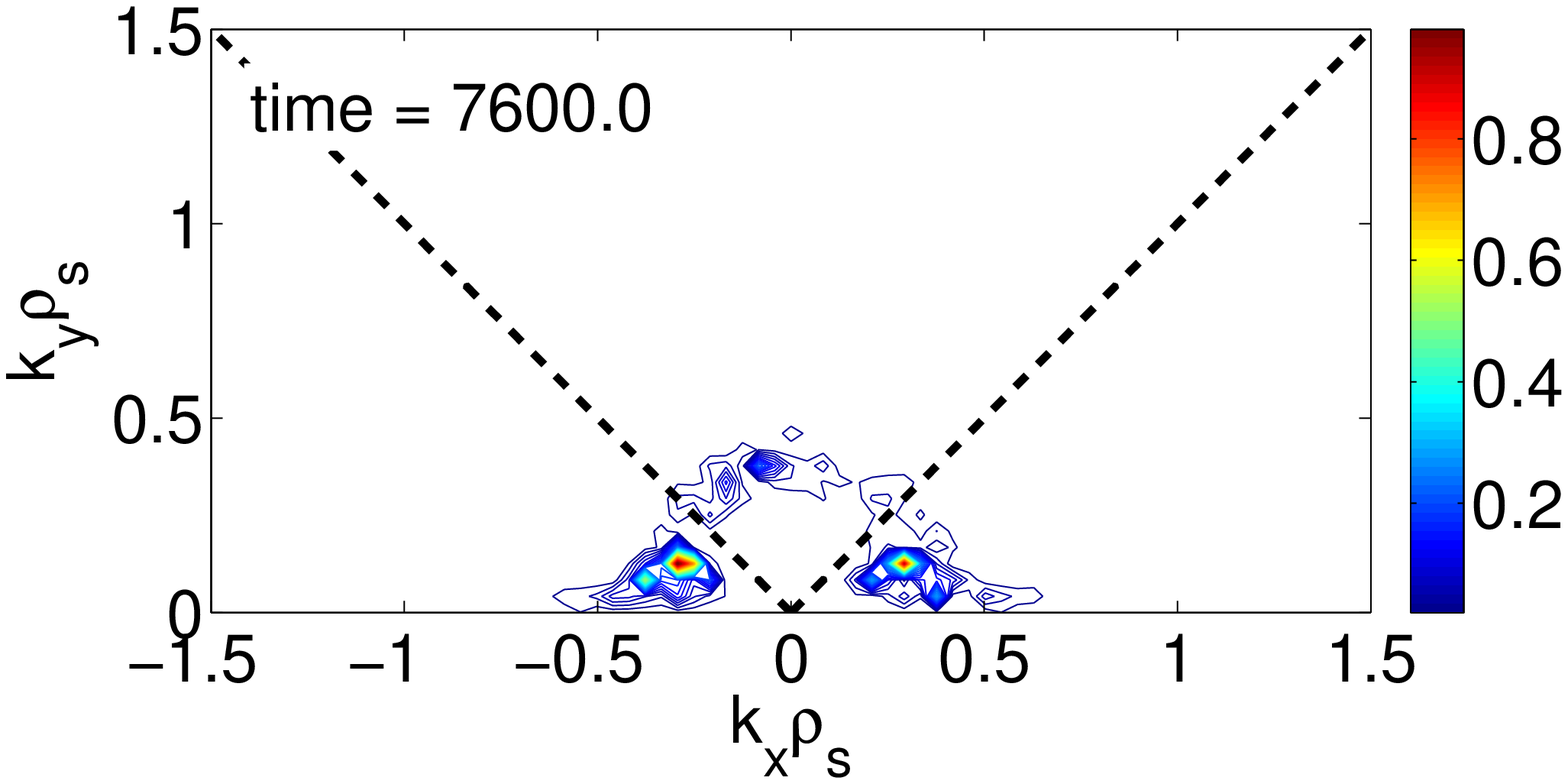} \\
    c) & d)
\end{tabular}
}
\caption{Snapshots of  the kinetic energy  spectrum normalized by its
maximum value. (a) $\mathscr{C}=1$, transition regime; (b) $\mathscr{C}=40$,
transition regime; (c) $\mathscr{C}=1$, saturated regime; (d) $\mathscr{C}=40$,
saturated regime.}
\label{ris:fld_denEk_c1_c40}
\end{figure}

\begin{figure}[h]
\includegraphics[width=7cm]
{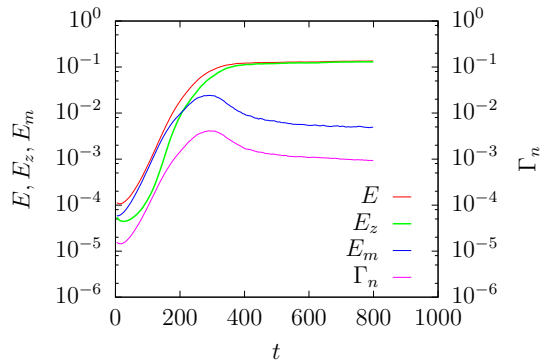}
\caption{Evolution of the total energy (red line), energy contained in a zonal
sector (green line), energy contained in a meridional sectors (blue line),
particle flux $\Gamma_{n}$ (magenta line). }
\label{ris:energy_1024_ck}
\end{figure}

For a better understanding of the anisotropic energy distributions,   on Fig.~\ref{ris:fld_denEk_c1_c40} we show the 2D kinetic energy spectra normalized by their maxima  for the cases $\mathscr{C}=1$ and $\mathscr{C}=40$. In the initial phase, for both cases, one can observe a concentration of the kinetic energy in the region corresponding to the characteristic scales of the  drift wave instability, see Figs.~\ref{ris:fld_denEk_c1_c40}(a) and~\ref{ris:fld_denEk_c1_c40}(b). Such a linear mechanism generates  energy mainly in the meridional sector. For the saturated state, one can observe the distinct features  of the energy distribution in the 2D $\bf k$-space. We can see that for large values of $\mathscr{C}$ there is a domination of concentration of the kinetic energy in the zonal sector, which absorbs energy from the meridional drift waves, see Fig.~\ref{ris:energy_1024_difC}(c). These computations for $\mathscr{C}\gg 1$ are extremely long. Even though in the limit we should obtain the dynamics governed by the CHM  equations \cite{hasegawa1978}, this may only be approached for very high value $\mathscr{C}$. Also the increase of $\mathscr{C}$ decreases the growth rate instability of the drift waves. Thus,  comparison with Connaughton \textit{et al}'s. simulation of CHM is not straightforward, since they artificially added a forcing term in order to mimic a HW-type instability and in the limit of $\mathscr{C}\rightarrow\infty$ the HW system tends to the unforced CHM system. For $\mathscr{C}=40$ the zonal flows are not yet very pronounced in the physical space visualization Fig.~\ref{ris:fld_ps_c1_c40}(b), but very clear in Fourier space. Indeed, while in  Fig.~\ref{ris:fld_denEk_c1_c40}(c) we see that the saturated 2D energy spectrum isotropic for $\mathscr{C}=1$,
on Fig.~\ref{ris:fld_denEk_c1_c40}(d) we can see that the spectrum is strongly anisotropic and mostly zonal for the $\mathscr{C}=40$   case.
\paragraph{Wavenumber dependent $\mathscr{C}.$}
Now we will consider the case when the parameter $\mathscr{C}$ is defined according to the relation~(\ref{eq:c1hw}) for $\rho_{s}\approx0.02$ and $\kappa=0.3491$. For the given parameters maximum value of $\mathscr{C}$ equal to 0.453. Note that this case has in common with the MHW model~\cite{numata2007} that the coupling term in Eq.~(\ref{eq:HW1}) and (\ref{eq:HW2}) is zero for the mode $k_{y}=0$. The numerical simulations were performed for $\nu=5\cdot10^{-5}$  with $1024^2$ modes, box size  $L=300$ and $\triangle t=10^{-4}$. Time evolution of the total, the  zonal and the  meridional kinetic energies, as well as the  particle flux $\Gamma_{n}$, are shown on Fig.~\ref{ris:energy_1024_ck}. We observe a similar picture as before in the simulation with large constant adiabaticity parameter $\mathscr{C}=40$. Namely, the total and the zonal energies grow monotonously until they reach saturation, whereas the meridional energy and the transport initially grow, reach maxima, and then get reduced so that their saturated levels are significantly less than their maximal values.  This is because the ZF's  draw energy from the drift waves, the same kind of LH-transition type process that we observed in the constant adiabaticity case with $\mathscr{C}=40$.
In the final saturated state,  there is a steady state of transfer of the energy from the drift waves to the ZF structures, so that the waves in the linear instability range in the meridional sector do not get a chance to grow, which can be interpreted as a nonlinear suppression of the drift-dissipative instability.

\begin{figure}[t]
\center{
\begin{tabular}{cc}
  \includegraphics[width=5.5cm]
{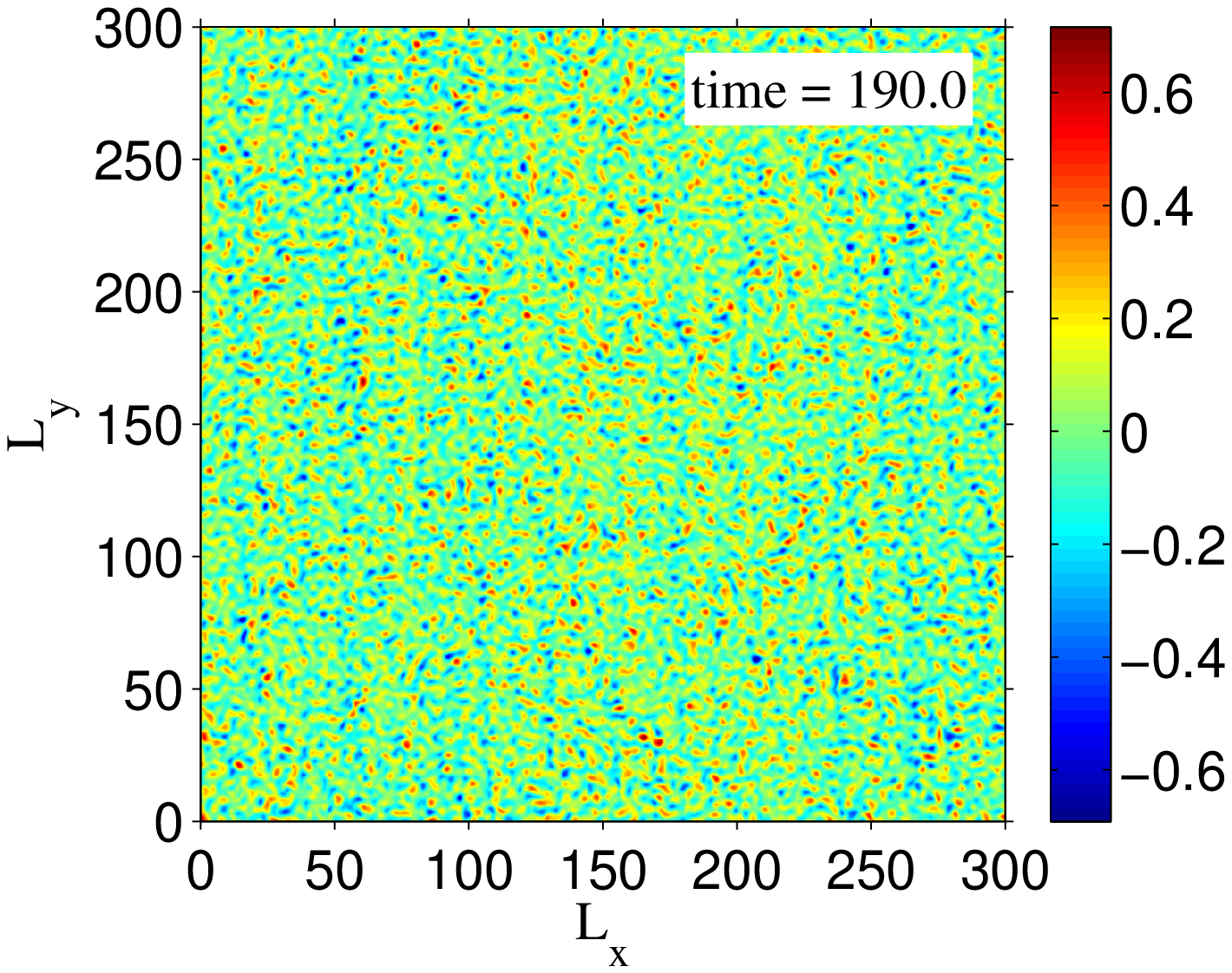} &
  \includegraphics[width=5.5cm]
{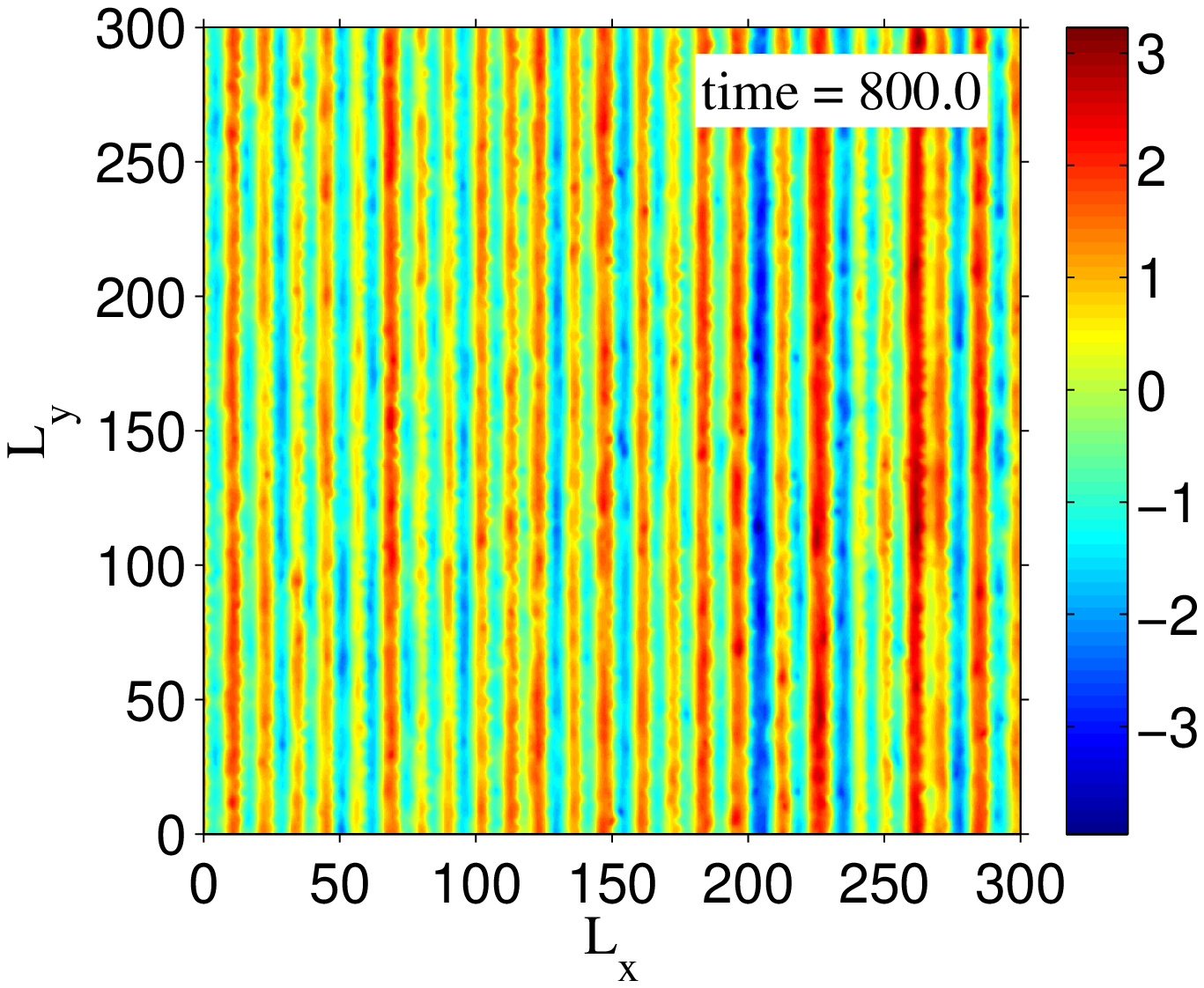} \\
    (a) & (b)
\end{tabular}
}
\caption{Streamfunction field at (a) $t=190$,
(b) $t=800$, for case $\mathscr{C}=\mathscr{C}(k)$.}
\label{ris:fld_ps_ck}
\end{figure}

\begin{figure}[t]
\center{
\begin{tabular}{ccc}
  \includegraphics[width=5.5cm]
{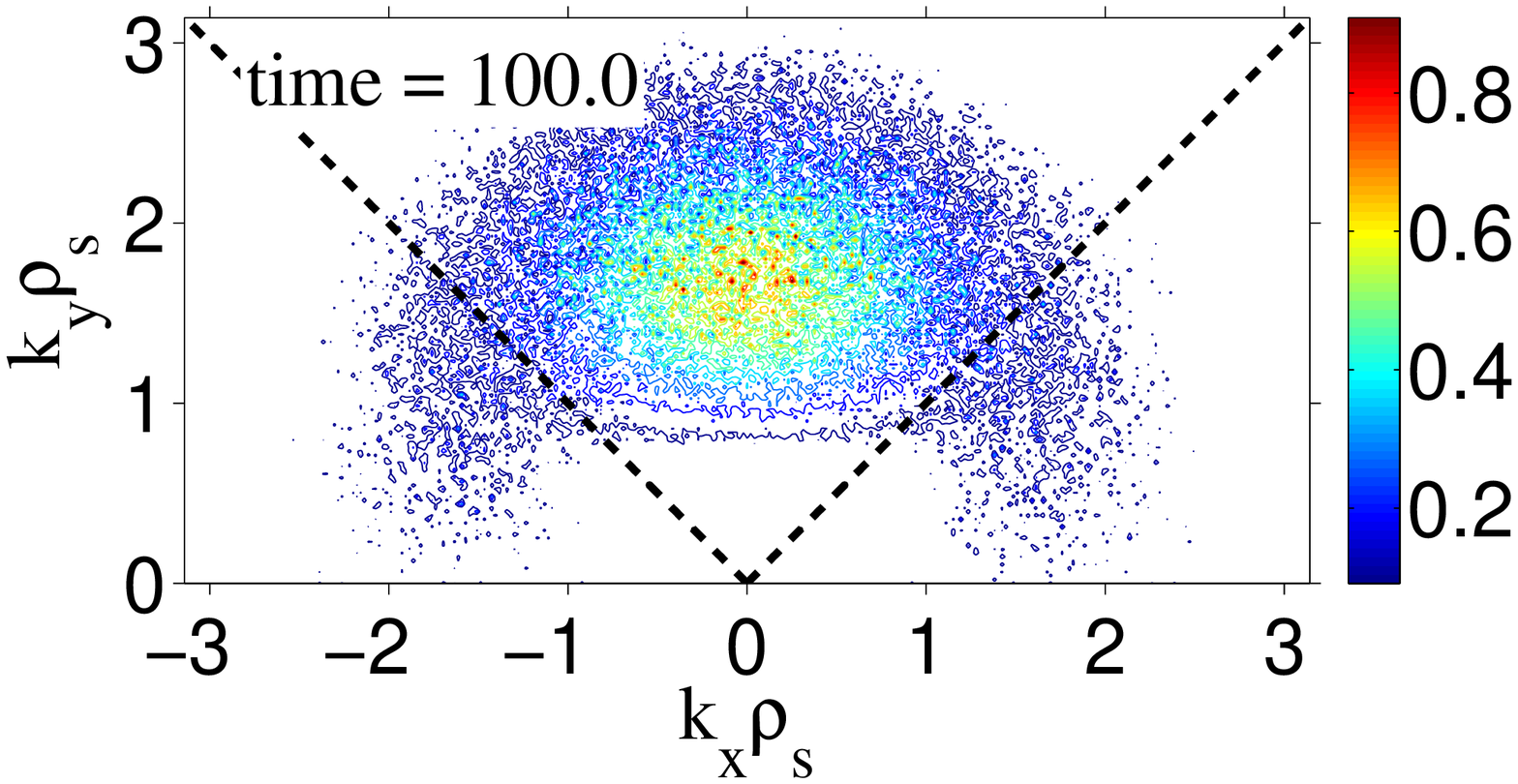} &
  \includegraphics[width=5.5cm]
{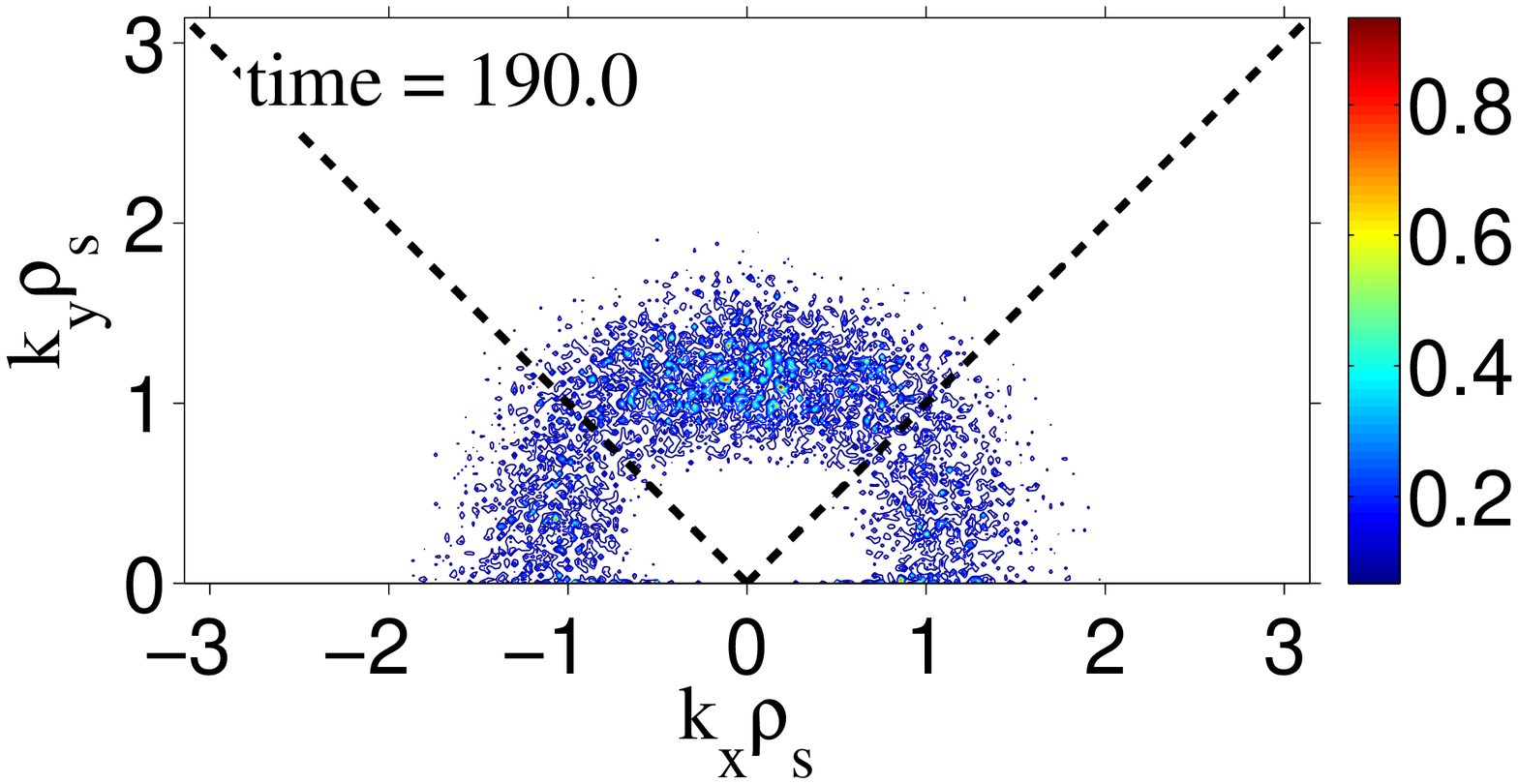} &
  \includegraphics[width=5.5cm]
{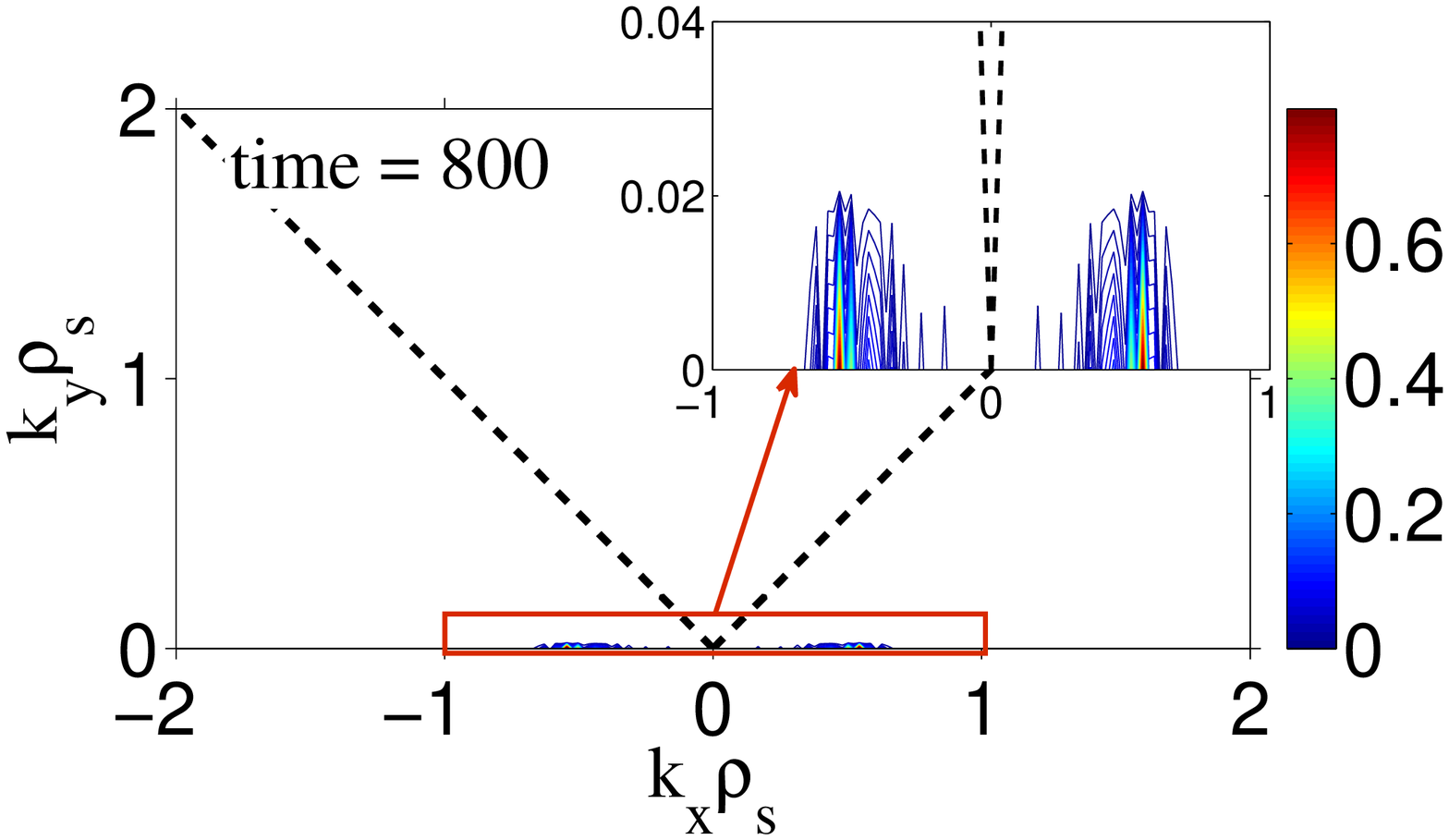} \\
    (a) & (b) & (c)
\end{tabular}
}
\caption{Snapshots of  the 2D kinetic energy spectrum normalized by its
maximum value and evaluated at (a) $t=100$,
(b) $t=190$ and (c) $t=800$ for the case of $\bf k$-dependent $\mathscr{C}$. For figure (c) a close-up view of a zonal sector is shown  in the right-hand corner.}
\label{ris:fld_denEk_ck}
\end{figure}

The physical space structure of the streamfunction $\psi$   is shown in Fig.~\ref{ris:fld_ps_ck}(a) and Fig.~\ref{ris:fld_ps_ck}(b) for an early moment and for the saturated state. One can see formation of well-formed ZF's.
Figs~\ref{ris:fld_denEk_ck}(a), (b) and (c) show the snapshots of the 2D energy spectrum evaluated at time $t=100$,  $t=190$ and $t=800$. We can see that initially meridional scales are excited via the linear instability mechanism, see Fig~\ref{ris:fld_denEk_ck}(a). This is followed by the nonlinear redistribution of the energy into the zonal sector, so that the spectrum for  $t=190$ looks almost isotropic,  Fig~\ref{ris:fld_denEk_ck}(b). The process of transfer to the zonal scales continues, and for $t=800$ we observe a very anisotropic spectrum which is mostly zonal, see Fig~\ref{ris:fld_denEk_ck}(c).

\section{Conclusions}
In this paper we have studied numerically turbulence described by the Hasegawa-Wakatani model Eq.~(\ref{eq:HW1}) and (\ref{eq:HW2}) for three different constant values of the adiabaticity parameter $\mathscr{C}$, and for a wavenumber-dependent $\mathscr{C}$ chosen to correspond to the fastest growing modes of the drift-dissipative instability. Our aim was to resolve a visible contradiction between the assertion made by Numata {\it et al.}\cite{numata2007} that zonal flows (ZF's) do not form in the original (unmodified) HW model and the clear observation of the ZF's by Connaughton {\it et al.}\cite{Nazarenko2011} within the Charney-Hasegawa-Mima model which is a limiting case of the HW system for large $\mathscr{C}$.

In our simulations for large values of $\mathscr{C}$, namely for  $\mathscr{C}=40$, we do observe formation of a strongly anisotropic flow, dominated by kinetic energy in the zonal sector, followed by the suppression of the short drift waves, drift-dissipative instability and the particle flux, as originally proposed in  the drift-wave/ZF feedback scenario put forward by Balk {\it et al.} in~1990 \cite{Balk1990a,Balk1990b}. This result suggests the original HW model to be the minimal nonlinear PDE model which can predict the LH transition. Note that even though the  drift-wave/ZF loop was also observed in the CHM simulations\cite{Nazarenko2011}, it cannot be considered a minimal model for the  LH transitions because the CHM model itself does not contain any instability, and it had to be mimicked by  an additional forcing term.

On the other hand, like in Numata {\it et al.}~\cite{numata2007}, we see neither formation of ZF's nor suppression of the short drift waves and the transport for low  values of  $\mathscr{C}$, 0.1 and 1. This is quite natural because in the limit of low $\mathscr{C}$ the HW model becomes a system similar to the isotropic 2D Navier-Stokes equation. Our guess is that Numata {\it et al.}~\cite{numata2007} have not explored the range of large $\mathscr{C}$ in their simulations, and therefore reached a conclusion that the original HW model would never produce ZF's. Note that the large $\mathscr{C}$ simulations are extremely demanding computationally because of the slow character of the linear instability in this case.

For the  simulation with wavenumber-dependent $\mathscr{C}$, with a maximum parameter $\mathscr{C}_{max}=0.453$, we have also observed the predicted drift-wave/ZF feedback process characterised by even stronger and pronounced zonation than in the $\mathscr{C}=40$ case. Of course, a decision which case is more relevant, constant or wavenumber-dependent $\mathscr{C}$, or the modified Hasegawa-Wakatani model~\cite{numata2007} should be decided based on the plasma parameters and the physical dimensions of the fusion device. Namely, the  wavenumber-dependent $\mathscr{C}$ should only be adopted if the fastest growing modes have wavenumbers allowed by the largest circumference of the tokamak; otherwise one should fix the parallel wavenumber at the lowest allowed value.

In this paper once again we have demonstrated the usefulness of the basic PDE models of plasma, like  CHM and HW, for predicting and describing important physical effects which are robust enough to show up in more realistic and less tractable plasma setups. Recall that the HW model is relevant for the tokamak edge plasma. As a step toward increased realism, one could consider a nonlinear three-field model  for the nonlinear ion-temperature gradient (ITG) instability system, which is relevant for the core plasma; see eg. Leboeuf {\it et al}~\cite{Leboeuf}. This model is somewhat more complicated than what we have considered so far but still tractable by similar methods. This is an important subject for future research.

\section{Acknowledgement}
Sergey Nazarenko gratefully acknowledges support from the government of Russian Federation  via grant No. 12.740.11.1430 for supporting research of teams working under supervision of invited scientists. Wouter Bos is supported by the contract SiCoMHD (ANR-Blanc 2011-045).

\bibliography{article_HW}

\begin{thebibliography}{13}%
\makeatletter
\providecommand \@ifxundefined [1]{%
 \@ifx{#1\undefined}
}%
\providecommand \@ifnum [1]{%
 \ifnum #1\expandafter \@firstoftwo
 \else \expandafter \@secondoftwo
 \fi
}%
\providecommand \@ifx [1]{%
 \ifx #1\expandafter \@firstoftwo
 \else \expandafter \@secondoftwo
 \fi
}%
\providecommand \natexlab [1]{#1}%
\providecommand \enquote  [1]{``#1''}%
\providecommand \bibnamefont  [1]{#1}%
\providecommand \bibfnamefont [1]{#1}%
\providecommand \citenamefont [1]{#1}%
\providecommand \href@noop [0]{\@secondoftwo}%
\providecommand \href [0]{\begingroup \@sanitize@url \@href}%
\providecommand \@href[1]{\@@startlink{#1}\@@href}%
\providecommand \@@href[1]{\endgroup#1\@@endlink}%
\providecommand \@sanitize@url [0]{\catcode `\\12\catcode `\$12\catcode
  `\&12\catcode `\#12\catcode `\^12\catcode `\_12\catcode `\%12\relax}%
\providecommand \@@startlink[1]{}%
\providecommand \@@endlink[0]{}%
\providecommand \url  [0]{\begingroup\@sanitize@url \@url }%
\providecommand \@url [1]{\endgroup\@href {#1}{\urlprefix }}%
\providecommand \urlprefix  [0]{URL }%
\providecommand \Eprint [0]{\href }%
\providecommand \doibase [0]{http://dx.doi.org/}%
\providecommand \selectlanguage [0]{\@gobble}%
\providecommand \bibinfo  [0]{\@secondoftwo}%
\providecommand \bibfield  [0]{\@secondoftwo}%
\providecommand \translation [1]{[#1]}%
\providecommand \BibitemOpen [0]{}%
\providecommand \bibitemStop [0]{}%
\providecommand \bibitemNoStop [0]{.\EOS\space}%
\providecommand \EOS [0]{\spacefactor3000\relax}%
\providecommand \BibitemShut  [1]{\csname bibitem#1\endcsname}%
\let\auto@bib@innerbib\@empty
\bibitem [{\citenamefont {Wagner}\ \emph {et~al.}(1982)\citenamefont {Wagner},
  \citenamefont {Becker}, \citenamefont {Behringer},\ and\ \citenamefont
  {Campbell}}]{Wagner1982}%
  \BibitemOpen
  \bibfield  {author} {\bibinfo {author} {\bibfnamefont {F.}~\bibnamefont
  {Wagner}}, \bibinfo {author} {\bibfnamefont {G.}~\bibnamefont {Becker}},
  \bibinfo {author} {\bibfnamefont {K.}~\bibnamefont {Behringer}}, \ and\
  \bibinfo {author} {\bibnamefont {Campbell}},\ }\bibfield  {title} {\enquote
  {\bibinfo {title} {Regime of improved confinement and high beta in
  neutral-beam-heated divertor discharges of the asdex tokamak},}\ }\href
  {\doibase 10.1103/PhysRevLett.49.1408} {\bibfield  {journal} {\bibinfo
  {journal} {Phys. Rev. Lett.}\ }\textbf {\bibinfo {volume} {49}},\ \bibinfo
  {pages} {1408--1412} (\bibinfo {year} {1982})}\BibitemShut {NoStop}%
\bibitem [{\citenamefont {Charney}(1948)}]{Charney1948}%
  \BibitemOpen
  \bibfield  {author} {\bibinfo {author} {\bibfnamefont {J.~G.}\ \bibnamefont
  {Charney}},\ }\bibfield  {title} {\enquote {\bibinfo {title} {On the scale of
  atmospheric motions},}\ }\href@noop {} {\bibfield  {journal} {\bibinfo
  {journal} {Geofys. Publ. Oslo}\ }\textbf {\bibinfo {volume} {17}},\ \bibinfo
  {pages} {1--17} (\bibinfo {year} {1948})}\BibitemShut {NoStop}%
\bibitem [{\citenamefont {Hasegawa}\ and\ \citenamefont
  {Mima}(1978)}]{hasegawa1978}%
  \BibitemOpen
  \bibfield  {author} {\bibinfo {author} {\bibfnamefont {A.}~\bibnamefont
  {Hasegawa}}\ and\ \bibinfo {author} {\bibfnamefont {K.}~\bibnamefont
  {Mima}},\ }\bibfield  {title} {\enquote {\bibinfo {title}
  {Pseudo-three-dimensional turbulence in magnetized nonuniform plasma},}\
  }\href {\doibase 10.1063/1.862083} {\bibfield  {journal} {\bibinfo  {journal}
  {Phys. Fluids}\ }\textbf {\bibinfo {volume} {21}},\ \bibinfo {pages} {87--92}
  (\bibinfo {year} {1978})}\BibitemShut {NoStop}%
\bibitem [{\citenamefont {Hasegawa}, \citenamefont {Maclennan},\ and\
  \citenamefont {Kodama}(1979)}]{hasegawa1979nonlinear}%
  \BibitemOpen
  \bibfield  {author} {\bibinfo {author} {\bibfnamefont {A.}~\bibnamefont
  {Hasegawa}}, \bibinfo {author} {\bibfnamefont {C.}~\bibnamefont {Maclennan}},
  \ and\ \bibinfo {author} {\bibfnamefont {Y.}~\bibnamefont {Kodama}},\
  }\bibfield  {title} {\enquote {\bibinfo {title} {Nonlinear behavior and
  turbulence spectra of drift waves and rossby waves},}\ }\href@noop {}
  {\bibfield  {journal} {\bibinfo  {journal} {Phys. Fluids}\ }\textbf {\bibinfo
  {volume} {22}},\ \bibinfo {pages} {2122} (\bibinfo {year}
  {1979})}\BibitemShut {NoStop}%
\bibitem [{\citenamefont {Rhines}(1975)}]{Rhines75}%
  \BibitemOpen
  \bibfield  {author} {\bibinfo {author} {\bibfnamefont {P.~B.}\ \bibnamefont
  {Rhines}},\ }\bibfield  {title} {\enquote {\bibinfo {title} {Waves and
  turbulence on a beta-plane.}}\ }\href@noop {} {\bibfield  {journal} {\bibinfo
   {journal} {J. Fluid Mech.}\ }\textbf {\bibinfo {volume} {69}},\ \bibinfo
  {pages} {417–443} (\bibinfo {year} {1975})}\BibitemShut {NoStop}%
\bibitem [{\citenamefont {Biglari}, \citenamefont {Diamond},\ and\
  \citenamefont {Terry}(1990)}]{biglari1990}%
  \BibitemOpen
  \bibfield  {author} {\bibinfo {author} {\bibfnamefont {H.}~\bibnamefont
  {Biglari}}, \bibinfo {author} {\bibfnamefont {P.~H.}\ \bibnamefont
  {Diamond}}, \ and\ \bibinfo {author} {\bibfnamefont {P.~W.}\ \bibnamefont
  {Terry}},\ }\bibfield  {title} {\enquote {\bibinfo {title} {Influence of
  sheared poloidal rotation on edge turbulence},}\ }\href {\doibase
  10.1063/1.859529} {\bibfield  {journal} {\bibinfo  {journal} {Phys. Fluids
  B}\ }\textbf {\bibinfo {volume} {2}},\ \bibinfo {pages} {1--4} (\bibinfo
  {year} {1990})}\BibitemShut {NoStop}%
\bibitem [{\citenamefont {Balk}, \citenamefont {Nazarenko},\ and\ \citenamefont
  {Zakharov}(1990{\natexlab{a}})}]{Balk1990a}%
  \BibitemOpen
  \bibfield  {author} {\bibinfo {author} {\bibfnamefont {A.}~\bibnamefont
  {Balk}}, \bibinfo {author} {\bibfnamefont {S.}~\bibnamefont {Nazarenko}}, \
  and\ \bibinfo {author} {\bibfnamefont {V.}~\bibnamefont {Zakharov}},\
  }\bibfield  {title} {\enquote {\bibinfo {title} {Nonlocal drift wave
  turbulence},}\ }\href@noop {} {\bibfield  {journal} {\bibinfo  {journal}
  {Sov. Phys. - JETP}\ }\textbf {\bibinfo {volume} {71}},\ \bibinfo {pages}
  {249--260} (\bibinfo {year} {1990}{\natexlab{a}})}\BibitemShut {NoStop}%
\bibitem [{\citenamefont {Balk}, \citenamefont {Nazarenko},\ and\ \citenamefont
  {Zakharov}(1990{\natexlab{b}})}]{Balk1990b}%
  \BibitemOpen
  \bibfield  {author} {\bibinfo {author} {\bibfnamefont {A.}~\bibnamefont
  {Balk}}, \bibinfo {author} {\bibfnamefont {S.}~\bibnamefont {Nazarenko}}, \
  and\ \bibinfo {author} {\bibfnamefont {V.}~\bibnamefont {Zakharov}},\
  }\bibfield  {title} {\enquote {\bibinfo {title} {On the nonlocal turbulence
  of drift type waves},}\ }\href {\doibase 10.1016/0375-9601(90)90168-N}
  {\bibfield  {journal} {\bibinfo  {journal} {Phys. Letters A}\ }\textbf
  {\bibinfo {volume} {146}},\ \bibinfo {pages} {217 -- 221} (\bibinfo {year}
  {1990}{\natexlab{b}})}\BibitemShut {NoStop}%
\bibitem [{\citenamefont {Dorland}\ and\ \citenamefont
  {Hammett}(1993)}]{Hammett1992}%
  \BibitemOpen
  \bibfield  {author} {\bibinfo {author} {\bibfnamefont {W.}~\bibnamefont
  {Dorland}}\ and\ \bibinfo {author} {\bibfnamefont {G.~W.}\ \bibnamefont
  {Hammett}},\ }\bibfield  {title} {\enquote {\bibinfo {title} {Gyrofluid
  turbulence models with kinetic effects},}\ }\href {\doibase 10.1063/1.860934}
  {\bibfield  {journal} {\bibinfo  {journal} {Phys. Fluids B}\ }\textbf
  {\bibinfo {volume} {5}},\ \bibinfo {pages} {812--835} (\bibinfo {year}
  {1993})}\BibitemShut {NoStop}%
\bibitem [{\citenamefont {Hasegawa}\ and\ \citenamefont
  {Wakatani}(1983)}]{Hasegawa1983}%
  \BibitemOpen
  \bibfield  {author} {\bibinfo {author} {\bibfnamefont {A.}~\bibnamefont
  {Hasegawa}}\ and\ \bibinfo {author} {\bibfnamefont {M.}~\bibnamefont
  {Wakatani}},\ }\bibfield  {title} {\enquote {\bibinfo {title} {Plasma edge
  turbulence},}\ }\href {\doibase 10.1103/PhysRevLett.50.682} {\bibfield
  {journal} {\bibinfo  {journal} {Phys. Rev. Lett.}\ }\textbf {\bibinfo
  {volume} {50}},\ \bibinfo {pages} {682--686} (\bibinfo {year}
  {1983})}\BibitemShut {NoStop}%
\bibitem [{\citenamefont {Numata}, \citenamefont {Ball},\ and\ \citenamefont
  {Dewar}(2007)}]{numata2007}%
  \BibitemOpen
  \bibfield  {author} {\bibinfo {author} {\bibfnamefont {R.}~\bibnamefont
  {Numata}}, \bibinfo {author} {\bibfnamefont {R.}~\bibnamefont {Ball}}, \ and\
  \bibinfo {author} {\bibfnamefont {R.~L.}\ \bibnamefont {Dewar}},\ }\bibfield
  {title} {\enquote {\bibinfo {title} {Bifurcation in electrostatic resistive
  drift wave turbulence},}\ }\href {\doibase 10.1063/1.2796106} {\bibfield
  {journal} {\bibinfo  {journal} {Phys. Plasmas}\ }\textbf {\bibinfo {volume}
  {14}},\ \bibinfo {eid} {102312} (\bibinfo {year} {2007})}\BibitemShut
  {NoStop}%
\bibitem [{\citenamefont {Connaughton}, \citenamefont {Nazarenko},\ and\
  \citenamefont {Quinn}(2011)}]{Nazarenko2011}%
  \BibitemOpen
  \bibfield  {author} {\bibinfo {author} {\bibfnamefont {C.}~\bibnamefont
  {Connaughton}}, \bibinfo {author} {\bibfnamefont {S.}~\bibnamefont
  {Nazarenko}}, \ and\ \bibinfo {author} {\bibfnamefont {B.}~\bibnamefont
  {Quinn}},\ }\bibfield  {title} {\enquote {\bibinfo {title} {Feedback of zonal
  flows on wave turbulence driven by small-scale instability in the
  charney-hasegawa-mima model},}\ }\href
  {http://stacks.iop.org/0295-5075/96/i=2/a=25001} {\bibfield  {journal}
  {\bibinfo  {journal} {Euro Phys. Lett.)}\ }\textbf {\bibinfo {volume} {96}},\
  \bibinfo {pages} {25001} (\bibinfo {year} {2011})}\BibitemShut {NoStop}%
\bibitem [{\citenamefont {Leboeuf}(2000)}]{Leboeuf}%
  \BibitemOpen
  \bibfield  {author} {\bibinfo {author} {\bibfnamefont {J.}~\bibnamefont
  {Leboeuf}},\ }\bibfield  {title} {\enquote {\bibinfo {title} {Full torus
  landau fluid calculations of ion temperature gradient-driven turbulence in
  cylindrical geometry},}\ }\href@noop {} {\bibfield  {journal} {\bibinfo
  {journal} {Phys. Plasmas}\ }\textbf {\bibinfo {volume} {7}},\ \bibinfo
  {pages} {5013} (\bibinfo {year} {2000})}\BibitemShut {NoStop}%
\end{thebibliography}%
\end{document}